\documentclass{scrarticle}

\usepackage{comment} 
\usepackage{physics}
\usepackage{hyperref}
\hypersetup{
    colorlinks=true,
    citecolor=blue,
    linkcolor=blue,      
    urlcolor=blue
    }

\usepackage{enumerate}
\usepackage{amsmath, amssymb}
\usepackage{graphicx,color}
\usepackage[
  backend=biber,
  style=numeric,
  sorting=none
]{biblatex}

\usepackage{authblk}

\makeatletter
\newcommand{\hrules}{%
\@for\i:=1,2,3,4,5\do{%
\noindent\red{\rule{\linewidth}{1pt}}\par\vspace{-1ex}%
}%
\vspace{1ex}
}
\makeatother

\begin{document}

\title{Machine-Learning-Empowered Quantum Sensing of the Plaquette Phase in a Three-Level Delta System}

\author[1]{Lorenzo Vitale}
\author[1]{Shreyasi Mukherjee}
\author[1,2]{Dario Fasone}
\author[1]{Enrico Martello}
\author[1,3]{Elisabetta Paladino}
\author[1,3]{Luigi Giannelli}
\author[1,3]{Giuseppe Falci}

\affil[1]{Dipartimento di Fisica e Astronomia ''Ettore Majorana'', Università di Catania, Italy}
\affil[2]{PhD in Quantum Technologies, Università di Napoli Federico II, Italy}
\affil[3]{INFN, Sezione di Catania, Italy}

\renewcommand\Authands{, }
\setlength{\affilsep}{0.5em}
\renewcommand\Affilfont{\small}

\date{}
\maketitle

\abstract{We propose a machine-learning-empowered approach to the quantum sensing of the plaquette phase, a gauge-invariant quantity arising in three-level $\Delta$ systems. This phase profoundly affects the system dynamics, breaking coherent population trapping and inducing a non-trivial phase dependence of the dynamics. 
We demonstrate that a multi-layer perceptron (MLP), trained in a supervised-learning framework, can accurately estimate the plaquette phase from STImulated Raman Adiabatic Passage (STIRAP) population transfer efficiencies measured under different driving conditions, which provide experimentally accessible observables. Our results highlight how the combination of coherent control and machine learning (ML) enables effective phase identification in closed-loop quantum systems, opening new perspectives for quantum technologies, specifically quantum sensing applications including synthetic gauge fields.}

\section{Introduction}

Over the last two decades, quantum technologies (QT) have undergone rapid development, driven by advances in the control and manipulation of individual quantum systems. By exploiting the main feature of quantum coherence, namely superposition and entanglement, as a fundamental resource, QT enable novel applications in quantum computation, communication, sensing, and metrology~\cite{acinQuantumTechnologiesRoadmap2018,dowlingQuantumTechnologySecond2003}. 
In this context, coherent control techniques are of paramount importance, as they enable both the implementation of quantum operations and the extraction of information from system dynamics~\cite{degenQuantumSensing2017}.

Among the simplest yet most paradigmatic platforms for investigating quantum control are three-level systems, which capture essential features of more complex architectures while remaining analytically tractable and experimentally accessible~\cite{shoreCoherentManipulationsAtoms2008}. In such systems, many control techniques exploit a quantum interference effect commonly referred to as coherent population trapping (CPT)~\cite{arimondoNonabsorbingAtomicCoherences1976,grayCoherentTrappingAtomic1978,arimondo1996v}. Stimulated Raman adiabatic passage (STIRAP) stands out as a widely used technique enabling robust and efficient population transfer between quantum states while suppressing occupation of lossy intermediate levels~\cite{bergmannCoherentPopulationTransfer1998,vitanovStimulatedRamanAdiabatic2017}.

Originally developed in atomic and molecular $\Lambda$-type systems, STIRAP has been successfully extended to solid-state platforms~\cite{kumarStimulatedRamanAdiabatic2016,siewertAdvancedControlCooperpair2009,falciDesignLambdaSystem2013,falciAdvancesQuantumControl2017,distefanoPopulationTransferLambda2015} and, more recently, to closed-loop $\Delta$-type configurations, where an additional coherent coupling closes the transition loop between the three levels~\cite{popeCoherentTrappingSmall2019}. While such configurations are typically forbidden in natural atomic systems due to electric-dipole selection rules~\cite{shoreCoherentManipulationsAtoms2008}, they can be realized in engineered platforms where these constraints are relaxed, such as superconducting artificial atoms~\cite{youAtomicPhysicsQuantum2011,liuOpticalSelectionRules2005a} or in mechanical metamaterials~\cite{velkovsky2024observation}.

In $\Delta$ systems, the closed-loop structure gives rise to a gauge-invariant phase associated with the triangular plaquette formed by the couplings. This plaquette phase profoundly affects the system dynamics by modifying interference effects and, in particular, by breaking the conditions for perfect coherent population trapping. As a result, population transfer protocols such as STIRAP acquire a non-trivial dependence on the phase, resulting in reduction of the transfer efficiency. Although this phase sensitivity degrades the performance of the protocol, it provides a valuable resource encoding information about the non-directly accessible phase into measurable observables.

This idea naturally connects to the broader framework of quantum sensing, where the intrinsic sensitivity of coherent quantum systems to external perturbations is harnessed to extract information with high precision~\cite{degenQuantumSensing2017}.

Reconstructing the plaquette phase from experimentally accessible observables constitutes a non-linear inverse problem, characterized by a highly non-trivial and generally non-invertible mapping between the phase and measurable quantities. This complexity suggests data-driven approaches capable of capturing hidden correlations in the system dynamics. 
In this context, ML provides a natural framework to approximate such mappings directly from data. 
ML-based techniques have demonstrated remarkable success across a wide range of quantum applications, including parameter estimation, identification of experimental configurations, and the development of optimized control strategies~\cite{marquardtMachineLearningQuantum2021,krennArtificialIntelligenceMachine2023,brown2021reinforcement,giannelliTutorialOptimalControl2022}.

Motivated by recent studies in which neural networks (NN) were successfully employed to diagnose global properties of noise in small quantum networks~\cite{mukherjeeNoiseClassificationThreelevel2024,fasoneDetectionNoiseCorrelations2025}, in this work we extend this approach to propose a ML-assisted sensing of the plaquette phase in a three-level $\Delta$ system. By exploiting the phase-dependent degradation of STIRAP transfer efficiency under different driving conditions, we train a NN to reconstruct the phase from experimentally accessible observables. This approach highlights how coherent control protocols, when combined with data-driven methods, can be repurposed as flexible and scalable tools for quantum sensing.

The paper is organized as follows. In Sec.~\ref{sec: System model}, we introduce the three-level $\Delta$ system and describe the emergence of the gauge phase and its impact on the adiabatic dynamics. In Sec.~\ref{sec: Methods}, we describe the data-driven framework employed to reconstruct the phase from experimentally accessible observables. In Sec.~\ref{sec: Results}, we present the training procedure and evaluate the performance of the regression model. Finally, Sec.~\ref{sec: Conclusions} summarizes the main conclusions and outlines possible future developments.

\section{System model}
\label{sec: System model}

\subsection{Closed-loop three-level systems} 

\begin{figure}
    \centering
    \includegraphics[width=0.4\linewidth]{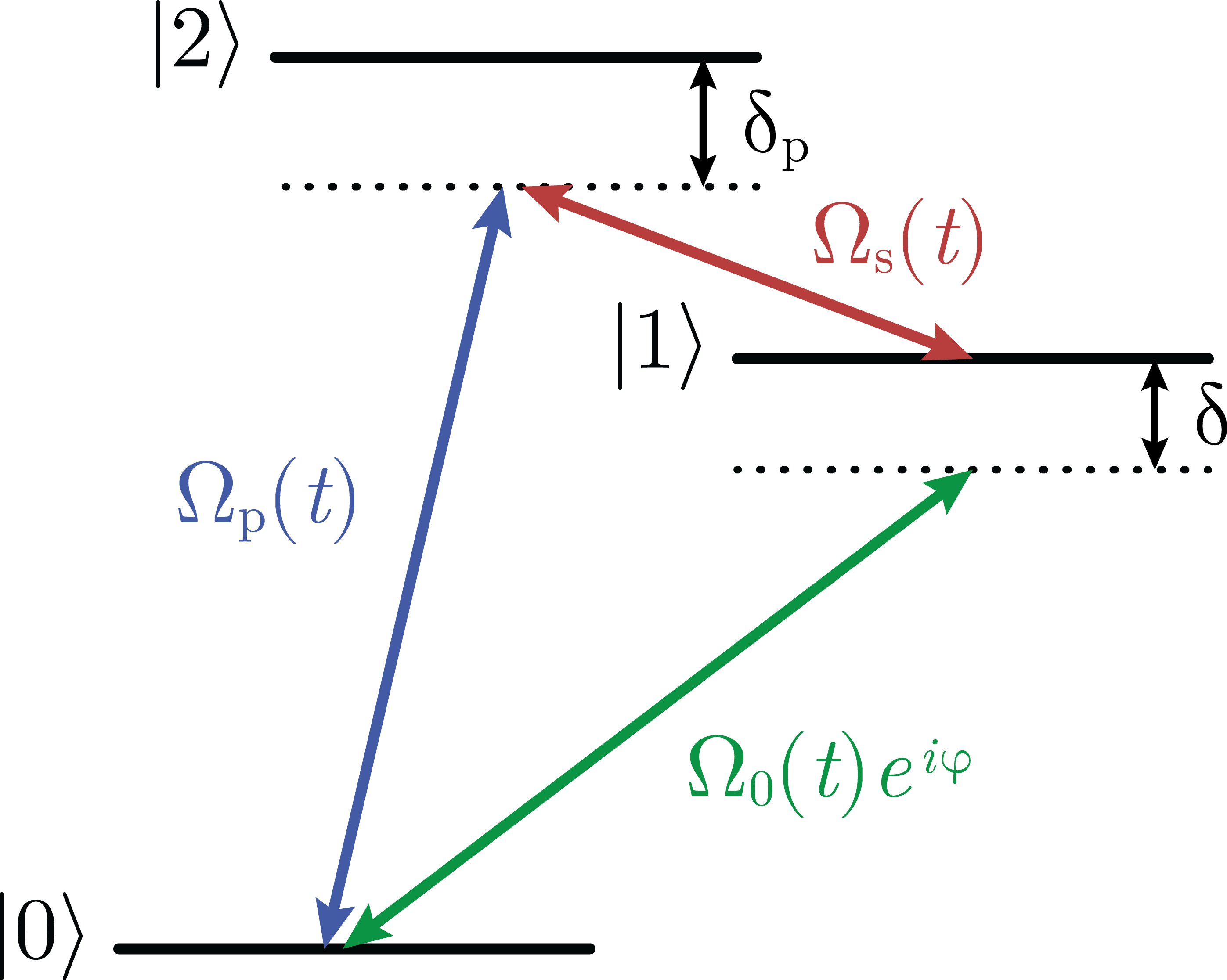}
    \caption{Three-level system in a $\Delta$ configuration.}
    \label{fig:three level delta system}
\end{figure}

We consider a $\Delta$-type system composed of three quantum states $\{\ket{0}, \ket{1}, \ket{2}\}$, 
mutually coupled to form a fully connected network (see Fig.~\ref{fig:three level delta system}). The closed-loop structure of the $\Delta$ configuration, corresponding to a triangular plaquette, is captured by the Hamiltonian ($\hbar=1$)~\cite{popeCoherentTrappingSmall2019} 
\begin{equation}
    H =
    \begin{pmatrix}
        0 & \Omega_0 & \Omega_p \\
        \Omega_0^\ast & \delta & \Omega_s \\
        \Omega_p^\ast & \Omega_s^\ast & \delta_p
    \end{pmatrix}.
    \label{equ:H_Delta_complex_couplings}
\end{equation}
In the language of quantum optics, the Hamiltonian in Eq.~\eqref{equ:H_Delta_complex_couplings} describes a multilevel atom driven by three nearly resonant laser fields in the rotating-wave approximation and expressed in a convenient multiple rotating frame. The diagonal terms represent detunings from resonance, while the off-diagonal elements are the Rabi frequencies associated to the amplitudes of the driving fields.
An equivalent description arises in a solid-state context, where the Hamiltonian corresponds to the tight-binding model of a network of three quantum dots arranged in a triangular geometry. In this picture, the diagonal elements $\delta$ and $\delta_p$ represent on-site energies (with respect to $\ket{0}$), while the couplings $\Omega_i$ describe tunable tunneling amplitudes between neighboring sites~\cite{popeCoherentTrappingSmall2019}.

Without loss of generality, the coupling elements $\Omega_p$ and $\Omega_s$ appearing in the Hamiltonian in Eq.~\eqref{equ:H_Delta_complex_couplings} can be taken to be real, since any phase factor can be absorbed into a redefinition of the basis states or, equivalently, by a unitary local phase transformation. However, it is noteworthy that this gauge freedom is not sufficient to make all couplings real simultaneously. As a consequence, one complex phase remains, which can be associated to the closed-loop structure of the $\Delta$ system~\cite{shoreCoherentManipulationsAtoms2008}.

In particular, choosing $\Omega_0$ to be real corresponds to fixing this residual phase to zero, a case discussed in details in Ref.~\cite{popeCoherentTrappingSmall2019}. In this work, we instead consider the general situation in which $\Omega_0$ is complex, thereby introducing a non-zero gauge phase that cannot be eliminated by local transformations and has direct physical consequences on the system dynamics.

\subsection{Emergence of a gauge phase}

The presence of a residual complex coupling, discussed above, can be formalized by explicitly identifying the gauge-invariant phase associated with the closed-loop structure.

In general, the couplings terms in Eq.~\eqref{equ:H_Delta_complex_couplings} are complex quantities which can be parametrized as $\Omega_k = |\Omega_k| e^{i\phi_k}$ with $k=0,p,s$. By applying a local phase transformation, $U=\mathrm{diag}(1,e^{-i(\phi_p-\phi_s)},e^{-i\phi_p})$, the Hamiltonian can be rewritten as
\begin{equation}
    H =
    \begin{pmatrix}
        0 & \Omega_0 e^{i\phi} & \Omega_p \\
        \Omega_0 e^{-i\phi} & \delta & \Omega_s \\
        \Omega_p & \Omega_s & \delta_p
    \end{pmatrix},
    \label{equ:H_Delta_phi}
\end{equation}
where all amplitudes are real and the residual phase $\phi = \phi_0 + \phi_s - \phi_p$ is a gauge-invariant quantity~\cite{shoreCoherentManipulationsAtoms2008}. This global phase corresponds to the total phase accumulated along the triangular plaquette. For this reason, we refer to it as the \emph{plaquette phase}.
Depending on the physical implementation of the $\Delta$ system, the plaquette phase can be interpreted either as an effective magnetic flux threading a triangular network of quantum dots~\cite{galitskiArtificialGaugeFields2019} or as the relative phase appearing in an atomic interferometer~\cite{buckleAtomicInterferometers1986a}.

Importantly, this phase is generally unknown and does not correspond to a directly measurable observable, but rather manifests itself through interference effects that affect the system dynamics~\cite{buckleAtomicInterferometers1986a}. As we will show in the following, its presence has profound consequences on the structure of the eigenstates and, in particular, on the existence and robustness of coherent population trapping. 
This observation naturally suggests the possibility of exploiting the resulting phase dependence of control protocols as an indirect sensing strategy.

\subsection{Coherent trapping and trapped state}
\label{sec: Coherent trapping and trapped state}

In three-level systems, quantum control protocols often exploit coherent population trapping, a quantum interference effect whereby two states, coherently coupled to a common intermediate level, combine to form a superposition state with no component in the intermediate state~\cite{arimondoNonabsorbingAtomicCoherences1976,grayCoherentTrappingAtomic1978,arimondo1996v}.
This superposition state, commonly referred to as a \emph{trapped state}, is effectively decoupled from the intermediate level, thereby confining the system dynamics to a reduced subspace. As a consequence, the intermediate state remains essentially unpopulated despite the presence of finite transition amplitudes. This mechanism lies at the core of population transfer protocols such as STIRAP, where adiabatic evolution of the system along the trapped state enables robust transfer between an initial and a target level while avoiding occupation of a typically lossy intermediate or excited state~\cite{bergmannCoherentPopulationTransfer1998,vitanovStimulatedRamanAdiabatic2017}.

Given the central role of the trapped state, we now investigate under which conditions the Hamiltonian~\eqref{equ:H_Delta_phi} admits an eigenstate of the form
\begin{equation}
    \ket{D} = c_0 \ket{0} + c_1 \ket{1},
    \label{equ:dark_state_structure}
\end{equation}
with no projection onto the level $\ket{2}$. By construction, such a state satisfies $\bra{2}\ket{D}=0$, implying that the system remains confined to the subspace spanned by $\{\ket{0},\ket{1}\}$ even in the presence of finite couplings to $\ket{2}$. In atomic physics, this decoupling suppresses fluorescence and motivates the denomination \emph{dark state}~\cite{shoreCoherentManipulationsAtoms2008}.

Following the treatment presented in Ref.~\cite{popeCoherentTrappingSmall2019}, for a three-level $\Delta$ system described by the Hamiltonian in Eq.~\eqref{equ:H_Delta_phi} the generalized trapping condition ensuring the existence of a dark state reads
\begin{equation}
    \delta = \frac{\Omega_0}{\Omega_p \Omega_s}
    \left( \Omega_s^2 e^{-i\phi} - \Omega_p^2 e^{i\phi} \right)
    \quad ; \quad
    z_D = -\frac{\Omega_p \Omega_0}{\Omega_s} e^{i\phi},
    \label{equ:exact_trapping_condition}
\end{equation}
where $z_D$ denotes the eigenvalue associated with the dark (trapped) state.
Under this condition, the corresponding eigenstate takes the form
\begin{equation}
    \ket{D} = \frac{\Omega_s \ket{0} - \Omega_p \ket{1}}{\Omega_{\mathrm{RMS}}}
    = \cos\theta \ket{0} - \sin\theta \ket{1},
    \label{equ:dark_state_delta}
\end{equation}
where $\Omega_{\mathrm{RMS}}=\sqrt{\Omega_p^2+\Omega_s^2}$ and the mixing angle $\theta$ is defined by $\tan\theta(t) = \Omega_p(t) / \Omega_s(t).$ 
The remaining two states form an effective Autler--Townes doublet with eigenvalues
\begin{equation}
    z_{\pm} = \frac{1}{2}
    \left[
        \delta_p + \frac{\Omega_s \Omega_0}{\Omega_p} e^{-i\phi}
        \pm \Omega_{\mathrm{AT}},
    \right],
\end{equation}
where 
\begin{equation}
    \Omega_{{AT}} =
    \sqrt{ \widetilde{\delta}_p^{\,2} + 4 \Omega_{\mathrm{RMS}}^2 },
    \qquad
    \widetilde{\delta}_p =
    \delta_p - \frac{\Omega_s \Omega_0}{\Omega_p} e^{-i\phi}.
\end{equation}
The corresponding eigenstates read:
\begin{equation}
    \ket{\pm}
    =
    \frac{2 i \Omega_0 \Omega_{\mathrm{RMS}} \sin\phi}{z_{\pm} - z_D}\, \ket{D}
    + \Omega_{\mathrm{RMS}} \ket{B}
    + \frac{1}{2}
    \left( \widetilde{\delta}_p \pm \Omega_{{AT}} \right) \ket{2},
    \label{equ:dressed_eigenstate_Delta_phi}
\end{equation}
where we defined the bright state $\ket{B} = \sin\theta\ket{0} + \cos\theta\ket{1}$, so that the set $\{\ket{D},\ket{B},\ket{2}\}$ forms a orthogonal basis.
For $\phi=0$ the above expressions reduce to the case of real-valued couplings, where the dressed states~\eqref{equ:dressed_eigenstate_Delta_phi} fully decouple from the dark eigenstate and coherent trapping is achieved~\cite{popeCoherentTrappingSmall2019}. The same behavior is recovered for $\phi=\pm\pi$, up to a sign inversion of the detuning $\delta$. 
For $\phi \neq 0,\pm\pi$, the dressed states, as shown in Eq.~\eqref{equ:dressed_eigenstate_Delta_phi}, acquire a finite projection onto the dark state, proportional to $\sin\phi$. 
This coupling is the origin of the breakdown of perfect coherent trapping in closed-loop configurations with a non-vanishing phase. As a consequence the adiabatic dynamics of the dark-state manifold is no longer protected, resulting in phase-dependent leakage. This suggests that the phase-dependent dynamics can be exploited as a resource for phase sensing.

\section{Methods}
\label{sec: Methods}

In this section, we introduce the sensing protocol and the data-driven framework used to reconstruct the plaquette phase. Leveraging on the phase sensitivity discussed above, we exploit the breakdown of coherent population trapping as a resource rather than a limitation. 

\subsection{Phase-sensitive STIRAP as a sensing protocol}
\label{sec:phase_sensing_protocol}

Enforcing the exact trapping condition~\eqref{equ:exact_trapping_condition} for $\phi \neq 0$ generally requires a complex detuning $\delta$, resulting in a non-Hermitian Hamiltonian. This corresponds to an idealized scenario in which gain and loss mechanisms are engineered to preserve an exact dark state. Furthermore, the required detuning depends explicitly on the plaquette phase, which is typically unknown and not directly controllable. In realistic settings, however, the detuning is real and phase-independent, and therefore perfect trapping cannot be achieved. As a consequence, deviations from ideal adiabatic transfer are unavoidable.

We adopt an operational approach by fixing the detuning to the value corresponding to perfect trapping, the zero-phase case for Eq.~\eqref{equ:zero_phase_trapping_condition}:
\begin{equation}
    \delta = \frac{\Omega_0}{\Omega_p \Omega_s}
    \left( \Omega_s^2 - \Omega_p^2 \right).
    \label{equ:zero_phase_trapping_condition}
\end{equation}
Using this as a baseline, we then investigate how the presence of a non-zero phase $\phi$ affects the STIRAP dynamics.

Population transfer from $\ket{0}$ to $\ket{1}$ is implemented via the STIRAP protocol as mentioned in Sec.~\ref{sec: Coherent trapping and trapped state}. In the standard $\Lambda$ configuration, complete population transfer is achieved at two-photon resonance, $\delta = 0$, by applying the so-called counterintuitive pulse sequence, in which the Stokes field $\Omega_s(t)$ precedes the pump field $\Omega_p(t)$~\cite{bergmannCoherentPopulationTransfer1998,vitanovStimulatedRamanAdiabatic2017}. 
A similar scheme can be extended to the three-level $\Delta$ system, following the treatment of Ref.~\cite{popeCoherentTrappingSmall2019}, where the effective two-photon detuning is governed by Eq.~\eqref{equ:zero_phase_trapping_condition}.

In this work we employ Gaussian envelopes of the form
\begin{equation}
    \Omega_p(t) = \Omega_p^{max} \exp\left[-\left(\frac{t-\tau}{T}\right)^2\right]
    \qquad;\quad
    \Omega_s(t) = \Omega_s^{max} \exp \left[-\left(\frac{t+\tau}{T}\right)^2\right],
    \label{equ1:gaussian_pulses}
\end{equation}
where $T$ denotes the pulse width, $\Omega^{max}_{p/s}$ the peak amplitude of the corresponding field, and $\tau$ the relative pulse delay. The counter-intuitive sequence requires $\tau>0$~\cite{giannelliTutorialOptimalControl2022}.

In the $\Delta$ configuration, the additional coupling $\Omega_0$ must also be taken into account. In order to implement a STIRAP-like protocol, this coupling is typically chosen to be non-zero only during the temporal overlap of the pump and Stokes pulses and to vanish asymptotically~\cite{popeCoherentTrappingSmall2019,unanyanLaserinducedAdiabaticAtomic1997}. A convenient choice satisfying these requirements is
\begin{equation}
    \Omega_0(t) = \frac{\Omega_s(t) \Omega_p(t) }{\Omega} 
                = \Omega e^{-2\frac{(t^2+\tau^2)}{T^2}},
    \label{equ:Omega0}
\end{equation}
where we have set $\Omega_p^{\max}=\Omega_s^{\max}=\Omega$.

\begin{figure}
    \centering
    \includegraphics[width=0.75\linewidth]{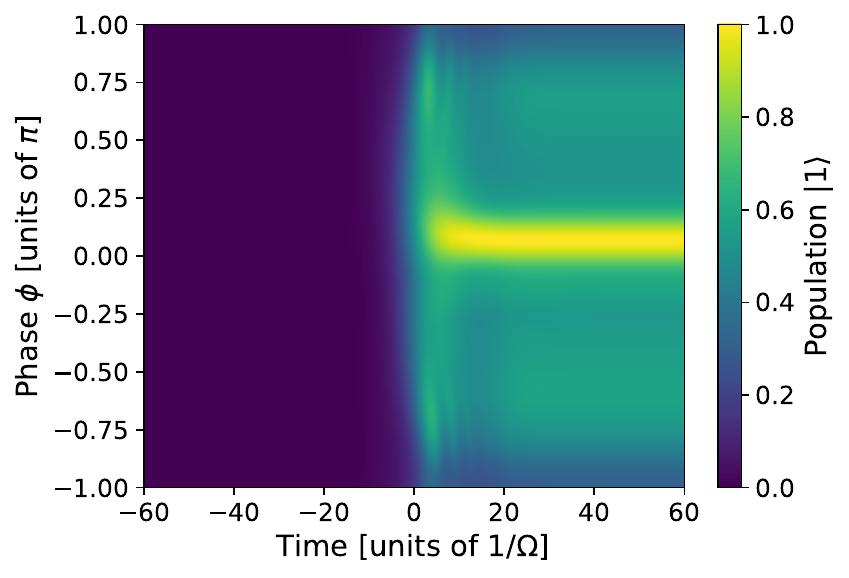}
    \caption{Time- and phase-resolved population of the target state $\ket{1}$ during the STIRAP protocol, obtained under the operational two-photon detuning condition of Eq.~\eqref{equ:zero_phase_trapping_condition}. The Gaussian pulses given by Eq.~\eqref{equ1:gaussian_pulses} are taken with equal peak amplitudes, $\Omega_p^{\max} = \Omega_s^{\max} = \Omega$. Simulation parameters: $\Omega T = 15$, $\tau = 0.7\,T$, and $\delta_p = 0$.}
    \label{fig:population_target_level}
\end{figure}

The presence of a non-zero plaquette phase prevents the exact fulfillment of the trapping condition~\eqref{equ:exact_trapping_condition} and leads to a breakdown of ideal adiabatic following. As a result, the population is no longer perfectly confined within the dark manifold and a residual leakage affects the transfer process. This effect manifests itself as a phase dependence of the population dynamics as shown in Fig.~\ref{fig:population_target_level}, where the occupation of the target state $\ket{1}$ varies with $\phi$.
In our protocol we propose to characterize this behavior only by measuring the final population of the target state (under different driving conditions), which we define as the transfer efficiency of the STIRAP protocol,
\begin{equation}
    \eta = |\braket{1}{\psi(t_\mathrm{f})}|^2.
    \label{equ:STIRAPeff}
\end{equation}
As shown in Fig.~\ref{fig:population_target_level}, the efficiency is maximal in the vicinity of $\phi = 0$ and decreases as the phase departs from this value due to leakage of the dark manifold. This phase dependence of the efficiency provides a simple and experimentally accessible observable, which will be exploited in the following as an input feature for a data-driven phase reconstruction scheme.

\subsection{Phase estimation with MLP}
To reconstruct the plaquette phase from experimentally accessible observables we adopt a supervised learning approach based on a feedforward NN, specifically a MLP. 
The network is trained on a labeled dataset, $\{(\mathbf{x}_i,\mathbf{\hat{y}}_i)\}_{i=1,\dots,N}$, where each input $\mathbf{x}_i$ is associated with a corresponding label $\mathbf{\hat{y}}_i$. 
The training consists in fitting a parametric function approximator (in this work the NN) by minimizing a cost function over the dataset, allowing the ML model to "learn" the mapping between inputs and outputs that generalizes to previously unseen data~\cite{marquardtMachineLearningQuantum2021,Burkov2019hundred}.

\begin{table}
    \centering
    \begin{tabular}{ccc}
    \hline
    Layer & Number of neurons & Activation function \\
    \hline
    Input & 8 & -- \\
    Hidden 1 & 64 & ReLU \\
    Hidden 2 & 64 & LeakyReLU \\
    Hidden 3 & 32 & LeakyReLU \\
    Hidden 4 & 16 & LeakyReLU \\
    Output & 2 & -- \\
    \hline
    \end{tabular}
    \caption{Architecture of the neural network used for the task of phase identification.}
    \label{tab:NN_architecture}
\end{table}

In the present context, the NN is trained to capture the non-trivial relation between transfer efficiencies under different control conditions and the underlying phase.
Specifically, the input $\mathbf{x}_i$ consists of an 8-dimensional vector representing the measured transfer efficiencies, whose detailed construction is discussed in Sec.~\ref{sec: Data generation}. The output targets is related to the plaquette phase $\phi$. Since the phase is a periodic quantity defined modulo $2\pi$, a direct estimation would introduce artificial discontinuities at the boundaries of the interval. To avoid this issue, the phase is encoded using a circular representation, and the NN is trained to predict the two-dimensional target vector
\begin{equation}
    \mathbf{\hat{y}}_i = \left(\cos \phi_i, \sin \phi_i \right),
\end{equation}
which provides a continuous and single-valued embedding of the phase value on the unit circle. 

The specific architecture implemented in this work is reported in Table~\ref{tab:NN_architecture}. The model is trained by minimizing the mean squared error loss,
\begin{equation}
    C \left( \{ \mathbf{y}_i, \hat{\mathbf{y}}_i \}_i  \right)
    = \frac{1}{N} \sum_{i=1}^N \left( \mathbf{y}_i - \hat{\mathbf{y}}_i \right)^2,
    \label{equ:cost_function}
\end{equation}
which measures the average quadratic deviation between the network output and the true target value.
The optimization of the network parameters is carried out using the adaptive variant of stochastic gradient descent Adam~\cite{kingma2014adam}. The gradients required for the parameter updates are efficiently computed via the backpropagation algorithm~\cite{marquardtMachineLearningQuantum2021,Burkov2019hundred}.

\subsection{Data generation}
\label{sec: Data generation}

The input to the neural network is a feature vector $\mathbf{x}_i \in \mathbb{R}^8$ composed of eight STIRAP transfer efficiencies evaluated under different driving and detuning conditions.
The use of multiple configurations is essential, as a single efficiency does not provide sufficient information to uniquely determine the plaquette phase. We take inspiration from previous works on noise and phase classification in multilevel systems~\cite{mukherjeeNoiseClassificationThreelevel2024,fasoneDetectionNoiseCorrelations2025}.

The efficiencies are evaluated for three different driving conditions, corresponding to distinct ratios of the pulse amplitudes $\Omega_{p/s}(t)$:
\begin{enumerate}[i)]
    \item $\Omega_p^{\max} = \Omega_s^{\max}$,
    \item $\Omega_p^{\max} = 2\Omega_s^{\max}$,
    \item $\Omega_p^{\max} = \Omega_s^{\max}$/2,
\end{enumerate}
In all cases, the root-mean-square Rabi frequency,
\begin{equation}
    \Omega_{\mathrm{RMS}} = \sqrt{(\Omega_p^{\max})^2 + (\Omega_s^{\max})^2},
\end{equation}
is kept fixed to $\Omega_{\mathrm{RMS}} = \sqrt{2} \Omega$ where $\Omega$ is here considered as unit of energy. This constraint ensures that all driving configurations share the same overall energy scale, allowing for meaningful comparison between different configurations~\cite{mukherjeeNoiseClassificationThreelevel2024,fasoneDetectionNoiseCorrelations2025}.

For each driving condition, we evaluate the STIRAP transfer efficiency under different detuning protocols. The resulting feature vector is composed of eight efficiencies obtained as follows.
The first three components correspond to the three driving conditions [(i)--(iii)] evaluated at zero two-photon detuning, $\delta=0$.
The next three components are obtained using the time-dependent detuning given by the trapping condition in Eq.~\eqref{equ:zero_phase_trapping_condition}, always evaluated assuming the symmetric driving condition, $\Omega_p^{\max}=\Omega_s^{\max}$, while the transfer efficiency is computed for each of the three driving configurations [(i)--(iii)].
Finally, the last two components are obtained by computing the time-dependent detuning, Eq.~\eqref{equ:zero_phase_trapping_condition}, using the same driving configuration employed to evaluate the transfer efficiency. This procedure is applied to the two asymmetric driving conditions [(ii) and (iii)], yielding two additional matched driving–detuning configurations.
Overall, this gives eight transfer efficiencies, which constitute the input feature vector. The corresponding driving--detuning combinations are summarized in Table~\ref{tab:detuning_strategies}.

\begin{table}
    \centering
    \begin{tabular}{c|cccc}
    \hline
    Driving & $\delta=0$ & $\delta_{\mathrm{sym}}$ & $\delta_{\mathrm{pump}}$ & $\delta_{\mathrm{Stokes}}$ \\
    \hline
    $\Omega_p^{\max}=\Omega_s^{\max}$ & $\checkmark$ & $\checkmark$ & -- & -- \\
    $\Omega_p^{\max}>\Omega_s^{\max}$ & $\checkmark$ & $\checkmark$ & $\checkmark$ & -- \\
    $\Omega_p^{\max}<\Omega_s^{\max}$ & $\checkmark$ & $\checkmark$ & -- & $\checkmark$ \\
    \hline
    \end{tabular}
    \caption{Driving--detuning configurations included in the dataset. Here, $\delta_{\mathrm{sym}}$ denotes the detuning obtained from Eq.~\eqref{equ:zero_phase_trapping_condition} evaluated for the symmetric driving condition (i), while $\delta_{\mathrm{pump}}$ and $\delta_{\mathrm{Stokes}}$ refer to the detuning evaluated for the pump-dominated (ii) and Stokes-dominated (iii) driving conditions, respectively. The symbol $\checkmark$ indicates that the corresponding configuration is included in the feature set.}
    \label{tab:detuning_strategies}
\end{table}

A synthetic labeled dataset is generated by numerically simulating the STIRAP dynamics of the three-level system. Gaussian pump and Stokes pulses, defined in Eq.~\eqref{equ1:gaussian_pulses}, are applied in the counter-intuitive sequence, while the coupling $\Omega_0(t)$ is defined as in Eq.~\eqref{equ:Omega0}.
For each configuration, the time-dependent Schr\"odinger equation associated with the Hamiltonian in Eq.~\eqref{equ:H_Delta_phi} is solved for $N=2000$ uniformly spaced values of the phase $\phi_i \in [-\pi,\pi]$. The resulting transfer efficiencies are collected into the feature vector $\mathbf{x}_i = (\eta_{i,1}, \eta_{i,2}, \dots, \eta_{i,8})$, where $\eta_{i,k}$ denotes the efficiency associated with the $k$-th configuration at phase $\phi_i$. The dataset is finally split into training, validation, and test sets with a ratio $0.6:0.2:0.2$.

\section{Results}
\label{sec: Results}

\begin{figure}
    \centering
    \includegraphics[width=0.65\linewidth]{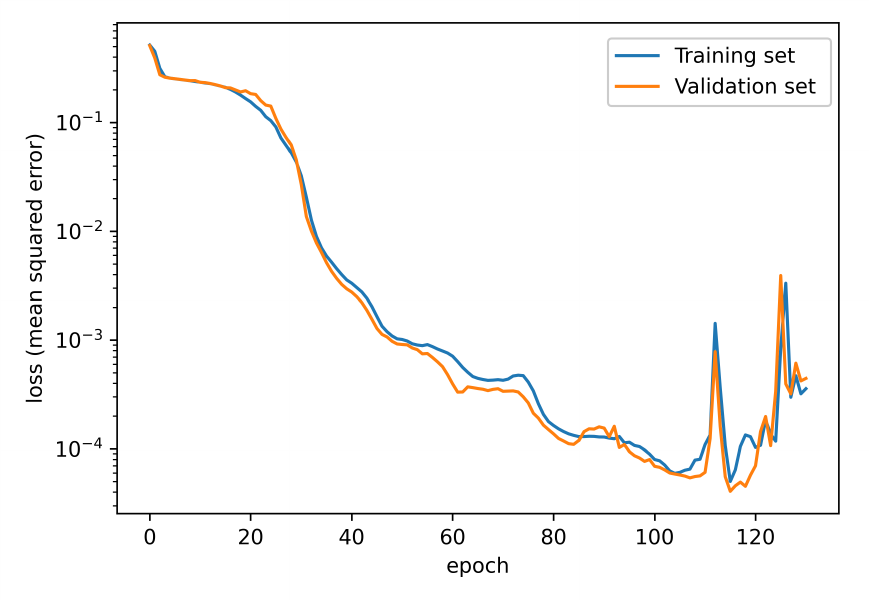}
    \caption{Training progress. The value of the cost function, given by Eq.~\eqref{equ:cost_function}, for the training sets (solid blue line) and the validation sets (solid orange line) vs. the number of epochs of training.}
    \label{fig:training}
\end{figure}

The MLP and the training procedure were implemented using the open-source Python library TensorFlow~\cite{tensorflow2015}. The network parameters, namely weights and biases, were initialized randomly prior to training, and subsequently optimized through gradient-based minimization of the loss function Eq.~\eqref{equ:cost_function}. 
The training process of the model is illustrated in Fig.~\ref{fig:training}. 
The cost function converges to small values, of the order of $10^{-3}$--$10^{-4}$.
Small fluctuations and occasional spikes are observed, due to the stochastic nature of the gradient-based optimization. Training is regularized via an early stopping criterion which halts the optimization once the validation loss ceases to improve. The absence of systematic deviations between training and validation losses indicates stable convergence without overfitting.

\begin{figure}
    \centering
    \includegraphics[width=\linewidth]{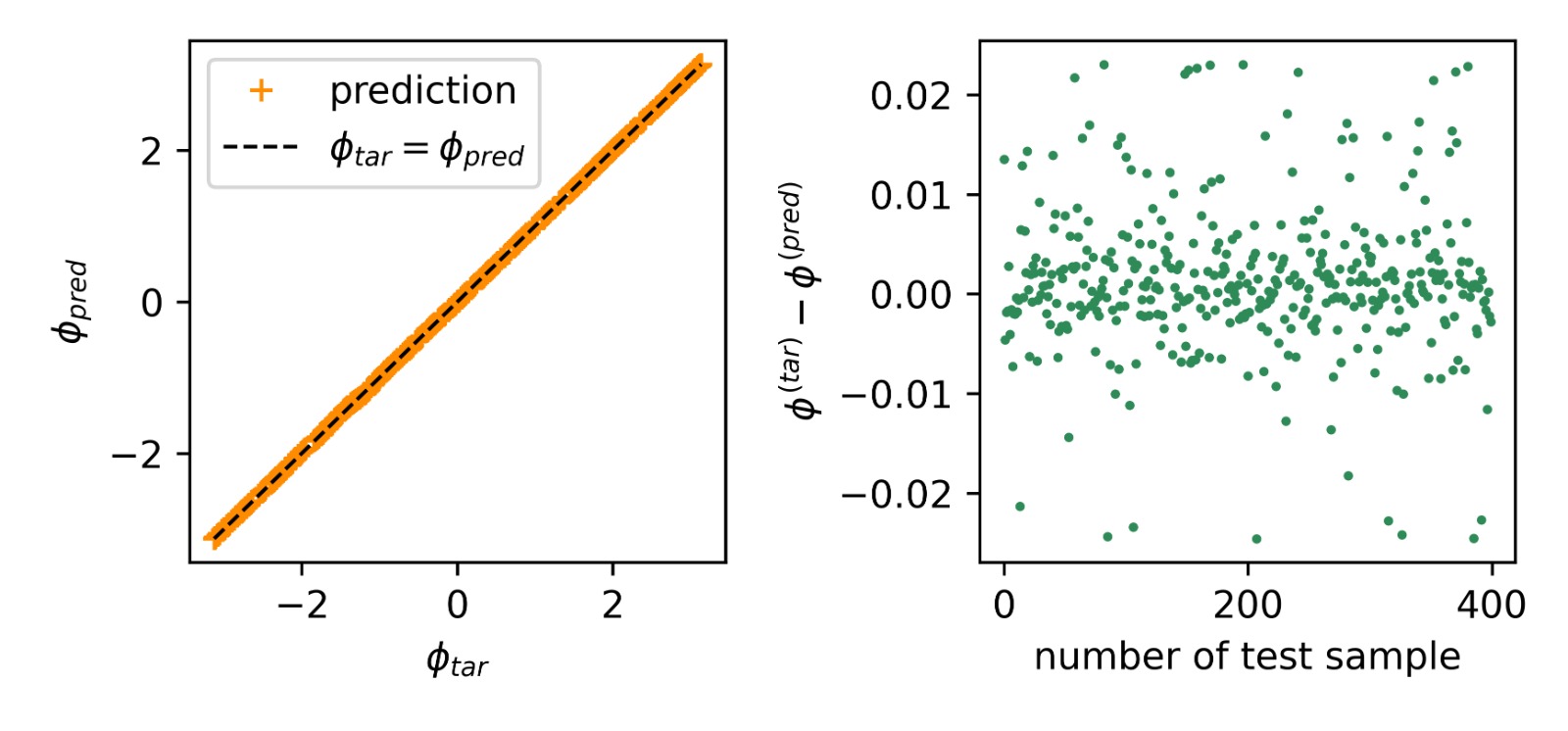}
    \caption{Performance of the neural network on the test dataset. \textbf{Left:} predicted phase versus the corresponding target value (the true value). \textbf{Right:} prediction accuracy as a function of the test sample index.}
    \label{fig:results_MLP}
\end{figure}

The performance of the trained model on previously unseen data (the test set) is summarized in Fig.~\ref{fig:results_MLP}. In the left panel, the predicted phase is plotted as a function of the corresponding target (true) phase for the test samples. The data points lie close to the identity line, demonstrating that the network is able to accurately reconstruct the plaquette phase from the transfer efficiencies.
Further insight is provided by the right panel of Fig.~\ref{fig:results_MLP}, where the prediction error, the deviation of the prediction from the target value, is shown as a function of the test sample index. The error remains consistently small across the dataset, indicating a high level of accuracy and robustness of the regression model. 
Taken together, these results demonstrate that the selected input features encode sufficient information to resolve the plaquette phase over the considered parameter range and that the trained neural network generalizes well beyond the training data.

\section{Conclusions}
\label{sec: Conclusions}

In this work, we have proposed a ML–assisted approach to the sensing of the plaquette phase in a three-level $\Delta$ system  with. By analyzing the impact of the gauge-invariant phase on the system dynamics, we have shown that the breakdown of coherent population trapping induces measurable deviations from ideal STIRAP transfer, which encode information about the phase.

From this observation, we have introduced a phase-sensitive STIRAP protocol in which transfer efficiencies, evaluated under different driving, are used as input features for a supervised learning model. Using synthetic data, we have demonstrated that the plaquette phase can be accurately reconstructed across the full interval range. The good generalization performance of the model indicates that the selected observables capture the essential phase-dependent signatures of the dynamics.

More generally, our results highlight how imperfections in coherent control protocols, often regarded as detrimental, can instead be harnessed as valuable resources for quantum sensing~\cite{fasoneDetectionNoiseCorrelations2025}. The combination of physically motivated control schemes with data-driven inference techniques provides a flexible framework for addressing nonlinear inverse problems in quantum systems.

The proposed approach is compatible with a variety of experimental platforms where $\Delta$-type configurations can be engineered, including superconducting circuits~\cite{liuOpticalSelectionRules2005a,roushanChiralGroundstateCurrents2017}, nitrogen–vacancy electronic spins~\cite{barfussPhasecontrolledCoherentDynamics2018} and quantum dots\cite{galitskiArtificialGaugeFields2019}.

Future developments may focus on reducing the number of input features required by the NN without significantly degrading its performance, as well as exploring alternative driving schemes beyond STIRAP that may enhance efficiency and experimental feasibility, and verifying the robustness of the protocol with respect to environmental noise.
These directions may further enhance the robustness and applicability of the method, paving the way toward practical implementations of phase-sensitive quantum sensing protocols in complex quantum devices

\section*{Acknowledgment}
E.M., L.G. and E.P. thank the PNRR MUR project PE0000023-NQSTI. 
E.P. thanks the COST Action CA21144 SuperQumap. 
G.F. thanks for the support ICSC - Centro Nazionale di Ricerca in High-Performance Computing, Big Data and Quantum Computing under project E63C22001000006, and Università degli Studi di Catania, project TCMQI PIACERI 2024/2026. 

\printbibliography

@article{acinQuantumTechnologiesRoadmap2018,
  title = {The Quantum Technologies Roadmap: A {{European}} Community View},
  shorttitle = {The Quantum Technologies Roadmap},
  author = {Ac{\'i}n, A. and Bloch, I. and Buhrman, H. and Calarco, T. and Eichler, C. and Eisert, J. and Esteve, D. and Gisin, N. and Glaser, S. J. and Jelezko, F. and Kuhr, S. and Lewenstein, M. and Riedel, M. F. and Schmidt, P. O. and Thew, R. and Wallraff, A. and Walmsley, I. and Wilhelm, F. K.},
  year = 2018,
  journal = {New Journal of Physics},
  volume = {20},
  pages = {080201},
  issn = {1367-2630},
  doi = {10.1088/1367-2630/aad1ea}
}

@article{dowlingQuantumTechnologySecond2003,
  title = {Quantum Technology: The Second Quantum Revolution},
  shorttitle = {Quantum Technology},
  author = {Dowling, J. P. and Milburn, G. J.},
  editor = {MacFarlane, A. G. J.},
  year = 2003,
  journal = {Philosophical Transactions of the Royal Society of London. Series A: Mathematical, Physical and Engineering Sciences},
  volume = {361},
  pages = {1655--1674},
  issn = {1364-503X, 1471-2962},
  doi = {10.1098/rsta.2003.1227}
}

@article{degenQuantumSensing2017,
  title = {Quantum Sensing},
  author = {Degen, C. L. and Reinhard, F. and Cappellaro, P.},
  year = 2017,
  journal = {Reviews of Modern Physics},
  volume = {89},
  pages = {035002},
  issn = {0034-6861, 1539-0756},
  doi = {10.1103/RevModPhys.89.035002}
}

@article{arimondoNonabsorbingAtomicCoherences1976,
  title = {Nonabsorbing Atomic Coherences by Coherent Two-Photon Transitions in a Three-Level Optical Pumping},
  author = {Arimondo, E. and Orriols, G.},
  year = 1976,
  journal = {Lettere al Nuovo Cimento (1971-1985)},
  volume = {17},
  number = {10},
  pages = {333--338},
  issn = {1827-613X},
  doi = {10.1007/BF02746514}
}

@article{grayCoherentTrappingAtomic1978,
  title = {Coherent Trapping of Atomic Populations},
  author = {Gray, H. R. and Whitley, R. M. and Stroud, C. R.},
  year = 1978,
  journal = {Optics Letters},
  volume = {3},
  number = {6},
  pages = {218--220},
  publisher = {Optica Publishing Group},
  issn = {1539-4794},
  doi = {10.1364/OL.3.000218}
}

@article{arimondo1996v,
  title = {V coherent population trapping in laser spectroscopy},
  author = {Arimondo, E.},
  journal = {Progress in optics},
  volume = {35},
  pages = {257--354},
  year = {1996},
  publisher = {Elsevier},
    doi = {10.1016/S0079-6638(08)70531-6},
}

@article{bergmannCoherentPopulationTransfer1998,
  title = {Coherent Population Transfer among Quantum States of Atoms and Molecules},
  author = {Bergmann, K. and Theuer, H. and Shore, B. W.},
  year = 1998,
  journal = {Reviews of Modern Physics},
  volume = {70},
  number = {3},
  pages = {1003--1025},
  publisher = {American Physical Society},
  doi = {10.1103/RevModPhys.70.1003}
}

@article{vitanovStimulatedRamanAdiabatic2017,
  title = {Stimulated {{Raman}} Adiabatic Passage in Physics, Chemistry, and Beyond},
  author = {Vitanov, N. V. and Rangelov, A. A. and Shore, B. W. and Bergmann, K.},
  year = 2017,
  journal = {Reviews of Modern Physics},
  volume = {89},
  number = {1},
  pages = {015006},
  publisher = {American Physical Society},
  doi = {10.1103/RevModPhys.89.015006}
}

@article{kumarStimulatedRamanAdiabatic2016,
  title = {Stimulated {{Raman}} Adiabatic Passage in a Three-Level Superconducting Circuit},
  author = {Kumar, K. S. and Veps{\"a}l{\"a}inen, A. and Danilin, S. and Paraoanu, G. S.},
  year = 2016,
  journal = {Nature Communications},
  volume = {7},
  number = {1},
  pages = {10628},
  publisher = {Nature Publishing Group},
  issn = {2041-1723},
  doi = {10.1038/ncomms10628}
}

@article{shoreCoherentManipulationsAtoms2008,
  title = {Coherent Manipulations of Atoms Using Laser Light},
  author = {Shore, B.},
  year = 2008,
  journal = {Acta Physica Slovaca. Reviews and Tutorials},
  volume = {58},
  number = {3},
  issn = {1336-040X, 0323-0465},
  doi = {10.2478/v10155-010-0090-z}
}

@article{galitskiArtificialGaugeFields2019,
  title = {Artificial Gauge Fields with Ultracold Atoms},
  author = {Galitski, V. and Juzeli{\=u}nas, G. and Spielman, I. B.},
  year = 2019,
  journal = {Physics Today},
  volume = {72},
  number = {1},
  pages = {38--44},
  issn = {0031-9228, 1945-0699},
  doi = {10.1063/PT.3.4111}
}

@article{buckleAtomicInterferometers1986a,
  title = {Atomic {{Interferometers}}: {{Phase-dependence}} in Multilevel Atomic Transitions},
  author = {Buckle, S. J. and Barnett, S. M. and Knight, P. L. and Lauder, M. A. and Pegg, D. T.},
  year = 1986,
  journal = {Optica Acta: International Journal of Optics},
  volume = {33},
  number = {9},
  pages = {1129--1140},
  issn = {0030-3909},
  doi = {10.1080/713822082}
}

@book{Burkov2019hundred,
  title = {The Hundred-Page Machine Learning Book},
  author = {Burkov, A.},
  year = {2019},
  volume = {1}
}

@article{marquardtMachineLearningQuantum2021,
  title = {Machine Learning and Quantum Devices},
  author = {Marquardt, F.},
  year = 2021,
  journal = {SciPost Physics Lecture Notes},
  pages = {29},
  issn = {2590-1990},
  doi = {10.21468/SciPostPhysLectNotes.29}
}

@misc{tensorflow2015,
title={ {TensorFlow}: Large-Scale Machine Learning on Heterogeneous Systems},
url={https://www.tensorflow.org/},
note={Software available from tensorflow.org},
author={
    M.~Abadi and
    A.~Agarwal and
    P.~Barham and
    E.~Brevdo and
    Z.~Chen and
    C.~Citro and
    G.~S.~Corrado and
    A.~Davis and
    J.~Dean and
    M.~Devin and
    S.~Ghemawat and
    I.~Goodfellow and
    A.~Harp and
    G.~Irving and
    M.~Isard and
    Y. Jia and
    R.~Jozefowicz and
    L.~Kaiser and
    M.~Kudlur and
    J.~Levenberg and
    D.~Man\'{e} and
    R.~Monga and
    S.~Moore and
    D.~Murray and
    C.~Olah and
    M.~Schuster and
    J.~Shlens and
    B.~Steiner and
    I.~Sutskever and
    K.~Talwar and
    P.~Tucker and
    V.~Vanhoucke and
    V.~Vasudevan and
    F.~Vi\'{e}gas and
    O.~Vinyals and
    P.~Warden and
    M.~Wattenberg and
    M.~Wicke and
    Y.~Yu and
    X.~Zheng},
  year={2015},
}

@article{youAtomicPhysicsQuantum2011,
  title = {Atomic Physics and Quantum Optics Using Superconducting Circuits},
  author = {You, J. Q. and Nori, F.},
  year = 2011,
  journal = {Nature},
  volume = {474},
  pages = {589--597},
  issn = {0028-0836, 1476-4687},
  doi = {10.1038/nature10122}
}

@article{liuOpticalSelectionRules2005a,
  title = {Optical {{Selection Rules}} and {{Phase-Dependent Adiabatic State Control}} in a {{Superconducting Quantum Circuit}}},
  author = {Liu, Y. and You, J. Q. and Wei, L. F. and Sun, C. P. and Nori, F.},
  year = 2005,
  journal = {Physical Review Letters},
  volume = {95},
  pages = {087001},
  publisher = {American Physical Society},
  doi = {10.1103/PhysRevLett.95.087001}
}

@article{unanyanLaserinducedAdiabaticAtomic1997,
  title = {Laser-Induced Adiabatic Atomic Reorientation with Control of Diabatic Losses},
  author = {Unanyan, R. G and Yatsenko, L. P and Bergmann, K. and Shore, B. W.},
  year = 1997,
  journal = {Optics Communications},
  volume = {139},
  pages = {48--54},
  issn = {00304018},
  doi = {10.1016/S0030-4018(97)00099-0}
}

@article{krennArtificialIntelligenceMachine2023,
  title = {Artificial intelligence and machine learning for quantum technologies},
  author = {Krenn, M. and Landgraf, J. and Foesel, T. and Marquardt, F.},
  journal = {Phys. Rev. A},
  volume = {107},
  issue = {1},
  pages = {010101},
  numpages = {22},
  year = {2023},
  publisher = {American Physical Society},
  doi = {10.1103/PhysRevA.107.010101}
}

@misc{kingma2014adam,
  title = {Adam: A Method for {{Stochastic Optimization}}},
  author = {Kingma, D. P. and Ba, J.},
  year = 2014,
  number = {arXiv:1412.6980},
  eprint = {1412.6980},
  primaryclass = {cs.LG},
  publisher = {arXiv},
  doi = {10.48550/arXiv.1412.6980},
  archiveprefix = {arXiv}
}

@article{roushanChiralGroundstateCurrents2017,
  title = {Chiral Ground-State Currents of Interacting Photons in a Synthetic Magnetic Field},
  author = {Roushan, P. and Neill, C. and Megrant, A. and Chen, Y. and Babbush, R. and Barends, R. and Campbell, B. and Chen, Z. and Chiaro, B. and Dunsworth, A. and Fowler, A. and Jeffrey, E. and Kelly, J. and Lucero, E. and Mutus, J. and O’Malley, P. J. J. and Neeley, M. and Quintana, C. and Sank, D. and Vainsencher, A. and Wenner, J. and White, T. and Kapit, E. and Neven, H. and Martinis, J.},
  date = {2017-02},
  journal = {Nature Physics},
  volume = {13},
  pages = {146--151},
  issn = {1745-2473, 1745-2481},
  doi = {10.1038/nphys3930}
}

@article{barfussPhasecontrolledCoherentDynamics2018,
  title = {Phase-Controlled Coherent Dynamics of a Single Spin under Closed-Contour Interaction},
  author = {Barfuss, A. and Kölbl, J. and Thiel, L. and Teissier, J. and Kasperczyk, M. and Maletinsky, P.},
  date = {2018-11},
  journal = {Nature Physics},
  shortjournal = {Nature Phys},
  volume = {14},
  pages = {1087--1091},
  publisher = {Nature Publishing Group},
  issn = {1745-2481},
  doi = {10.1038/s41567-018-0231-8},
}

@article{popeCoherentTrappingSmall2019,
  title = {Coherent Trapping in Small Quantum Networks},
  author = {Pope, T. J. and Rajendran, J. and Ridolfo, A. and Paladino, E. and Pellegrino, F. M. D. and Falci, G.},
  year = 2019,
  journal = {Journal of Statistical Mechanics: Theory and Experiment},
  volume = {2019},
  number = {12},
  pages = {124024},
  issn = {1742-5468},
  doi = {10.1088/1742-5468/ab54b7}
}

@article{siewertAdvancedControlCooperpair2009,
  title = {Advanced Control with a {{Cooper-pair}} Box: {{Stimulated Raman}} Adiabatic Passage and {{Fock-state}} Generation in a Nanomechanical Resonator},
  shorttitle = {Advanced Control with a {{Cooper-pair}} Box},
  author = {Siewert, J. and Brandes, T. and Falci, G.},
  year = 2009,
  journal = {Physical Review B},
  volume = {79},
  number = {2},
  pages = {024504},
  publisher = {American Physical Society},
  doi = {10.1103/PhysRevB.79.024504}
}

@article{falciDesignLambdaSystem2013,
  title = {Design of a {{Lambda}} System for Population Transfer in Superconducting Nanocircuits},
  author = {Falci, G. and La Cognata, A. and Berritta, M. and D'Arrigo, A. and Paladino, E. and Spagnolo, B.},
  year = 2013,
  journal = {Physical Review B},
  volume = {87},
  number = {21},
  pages = {214515},
  publisher = {American Physical Society},
  doi = {10.1103/PhysRevB.87.214515}
}

@article{falciAdvancesQuantumControl2017,
  title = {Advances in Quantum Control of Three-Level Superconducting Circuit Architectures},
  author = {Falci, G. and Di Stefano, P. G. and Ridolfo, A. and D'Arrigo, A. and Paraoanu, G. S. and Paladino, E.},
  year = 2017,
  journal = {Fortschritte der Physik},
  volume = {65},
  number = {6-8},
  pages = {1600077},
  issn = {1521-3978},
  doi = {10.1002/prop.201600077}
}

@article{distefanoPopulationTransferLambda2015,
  title = {Population Transfer in a {{Lambda}} System Induced by Detunings},
  author = {Di Stefano, P. G. and Paladino, E. and D'Arrigo, A. and Falci, G.},
  year = 2015,
  journal = {Physical Review B},
  volume = {91},
  number = {22},
  pages = {224506},
  publisher = {American Physical Society},
  doi = {10.1103/PhysRevB.91.224506}
}

@article{mukherjeeNoiseClassificationThreelevel2024,
  title = {Noise Classification in Three-Level Quantum Networks by {{Machine Learning}}},
  author = {Mukherjee, S. and Penna, D. and Cirinn{\`a}, F. and Paternostro, M. and Paladino, E. and Falci, G. and Giannelli, L.},
  year = 2024,
  journal = {Machine Learning: Science and Technology},
  volume = {5},
  number = {4},
  pages = {045049},
  issn = {2632-2153},
  doi = {10.1088/2632-2153/ad9193}
}

@article{giannelliTutorialOptimalControl2022,
  title = {A Tutorial on Optimal Control and Reinforcement Learning Methods for Quantum Technologies},
  author = {Giannelli, L. and Sgroi, S. and Brown, J. and Paraoanu, G. S. and Paternostro, M. and Paladino, E. and Falci, G.},
  year = 2022,
  journal = {Physics Letters A},
  volume = {434},
  pages = {128054},
  issn = {03759601},
  doi = {10.1016/j.physleta.2022.128054}
}

@misc{fasoneDetectionNoiseCorrelations2025,
  title = {Detection of Noise Correlations in Two Qubit Systems by {{Machine Learning}}},
  author = {Fasone, D. and Mukherjee, S. and Penna, D. and Cirinn{\`a}, F. and Paternostro, M. and Paladino, E. and Giannelli, L. and Falci, G. A.},
  year = 2025,
  number = {arXiv:2509.03389},
  eprint = {2509.03389},
  primaryclass = {quant-ph},
  publisher = {arXiv},
  doi = {10.48550/arXiv.2509.03389},
  archiveprefix = {arXiv}
}

@article{brown2021reinforcement,
  title={Reinforcement learning-enhanced protocols for coherent population-transfer in three-level quantum systems},
  author={Brown, J. and Sgroi, S. and Giannelli, L. and Paraoanu, G. S. and Paladino, E. and Falci, G. and Paternostro, M. and Ferraro, A.},
  journal={New Journal of Physics},
  volume={23},
  number={9},
  pages={093035},
  year={2021},
  publisher={IOP Publishing},
  doi = {10.1088/1367-2630/ac2393}
}

@article{velkovsky2024observation,
  title={Observation of chiral solitary waves in a nonlinear Aharonov-Bohm ring},
  author={Velkovsky, I. and Abraham, A. and Martello, E. and Yu, J. and Singhal, Y. and Gonzalez, A. and Lewis, D. and Price, H. and Ozawa, T. and Gadway, B.},
  journal={arXiv preprint arXiv:2406.01732},
  year={2024}
}

\end{document}